\input{aipcheck}

\documentclass[
    ,final            
  ]
  {aipproc}

\layoutstyle{6x9}

\begin{document}

\title{Measurements of Double-Spin Asymmetries in SIDIS of Longitudinally Polarized Leptons off 
Transversely Polarized Protons}

\classification{13.60.-r, 13.85.Ni, 13.87.Fh, 13.88.+e}
\keywords{Deep inelastic scattering, transverse momentum dependent distribution functions}

\author{L.L. Pappalardo}{
  address={INFN -- University of Ferrara - Dipartimento di Fisica, Via Saragat 1, 44100 Ferrara, Italy}
}

\author{M. Diefenthaler}{
  address={University of Illinois, Department of Physics, 1110 West Green Street, Urbana, USA\\
{\bf (on behalf of the HERMES Collaboration)}}
}

\begin{abstract}
A Fourier analysis of double-spin azimuthal asymmetries measured at HERMES in 
semi-inclusive deep-inelastic scattering of longitudinally polarized leptons off tranversely 
polarized protons is presented for pions and charged kaons. The extracted amplitudes can be 
interpreted as convolutions of transverse momentum-dependent distribution and fragmentation 
functions and provide sensitivity to e.g. the poorly known worm-gear quark distribution 
$g_{1T}^\perp$.
\end{abstract}

\maketitle


\section{accessing TMDs in Semi-inclusive DIS}

\indent
In recent years, semi-inclusive deep-inelastic-scattering (SIDIS) processes are being explored 
by several experiments to investigate the nucleon structure through the measurements of new 
observables, not accesible in inclusive DIS. The detection of a final-state hadron in coincidence 
with the scatterd lepton has the advantage of providing unique information on the quark 
flavors involved in the scattering process ("flavor tagging") through the identification of the 
final state hadrons (e.g. $\pi$, $K$, etc), and allows to access new dimensions, such as the 
transverse-spin and transverse-momentum degrees of freedom of the nucleon. For instance, the recent 
first extraction of the chiral-odd transversity distribution $h_1^q(x)$~\cite{ref:Ans07}, the least 
known of the three fundamental leading-twist collinear parton distribution functions (PDFs), 
 required the measurement of specific azimuthal asymmetries (the "Collins asymmetries") in SIDIS 
of unpolarized leptons off transversely polarized protons~\cite{ref:Air05,ref:Air10,ref:COMP10_pro} 
and deuterons~\cite{ref:COMP05_deu,ref:COMP07_deu}. 
Here $x$ denotes the fraction of the longitudinal momentum of the parent (fast-moving) nucleon 
carried by the active quark.

When the transverse momentum ${\bf p}_T$ of the quarks is not integrated out, a variety of new PDFs 
arise, describing correlations between the quark or the nucleon spin with the quark transverse 
momentum, often referred to as {\it spin-orbit correlations}. These poorly known PDFs, typically 
denoted as transverse-momentum-dependent PDFs (or simply TMDs), encode information on the 
3-dimensional structure of nucleons and are increasingly gaining theoretical and experimental 
interest. At leading-twist, eight TMDs, each with a specific probabilistic interpretation in terms 
of quark number densities, enter the SIDIS cross section in conjunction with a fragmentation function 
(FF) (see e.g.~\cite{ref:Bac07}). When the polarization of the final hadrons is not regarded, this can 
be either the chiral-odd Collins function $H_1^\perp(z,{\bf K}_T^2)$, describing left-right 
asymmetries in the fragmentation of transversely polarized quarks, or the relatively well known 
spin-independent chiral-even $D_{1}(z,{\bf K}_T^2)$ FF. Here $z$ and 
${\bf K}_T$ denote the fraction of the energy of the exchanged virtual photon carried by the produced 
hadron and the transverse momentum of the fragmenting quark with respect to the outgoing hadron 
direction, respectively.   

Among the leading-twist TMDs, the {\it 'worm-gear'} $h_{1L}^\perp(x,{\bf p}_T^2)$ and 
$g_{1T}^\perp(x,{\bf p}_T^2)$ are those that have received the least attention so far.
They are, nevertheless, very intriguing objects: $g_{1T}^\perp(x,{\bf p}_T^2)$ 
($h_{1L}^\perp(x,{\bf p}_T^2)$)
describes the probability of finding a longitudinally (transversely) polarized quark inside a 
transversely (longitudinally) polarized nucleon. Interestingly, they are the only two leading-twist 
TMDs whose corresponding Generalized Parton Distributions vanish in light-come quark 
models~\cite{ref:Die05}, and are found to be one the opposite of the other 
($g_{1T}^\perp(x,{\bf p}_T^2)=-h_{1L}^\perp(x,{\bf p}_T^2)$) in various quark
 models~\cite{ref:Jak97,ref:Pasq08,ref:Bac08,ref:Ava10}. Despite their similarities, these two TMDs 
have a different behaviour under chiral transformations: $h_{1L}^\perp(x,{\bf p}_T^2)$ is chiral-odd 
and can be probed in SIDIS in combination with the Collins FF, while 
$g_{1T}^\perp(x,{\bf p}_T^2)$ is chiral-even and can thus be accessed in SIDIS combined with the 
unpolarized FF. Another important difference, especially from the experimental 
point of view, is that $h_{1L}^\perp(x,{\bf p}_T^2)$ can be accessed in longitudinal target
$A_{UL}$ single-spin asymmetries (SSAs), whereas in the case of 
$g_{1T}^\perp(x,{\bf p}_T^2)$ the longitudinal polarization of the active quark leads to 
$A_{LT}$ double-spin asymmetries (DSAs), requiring both a longitudinally polarized beam and a 
transversely polarized target~\cite{ref:Bac07}. 

At leading-twist, the term of the SIDIS cross section that accounts for this DSA exhibits a 
${\rm cos}(\phi-\phi_S)$ modulation in the azimuthal angles $\phi$ and $\phi_S$, respectively of the 
detected hadron and of the target transverse polarization with respect to the lepton scattering plane 
and about the virtual-photon direction. In SIDIS experiments the worm-gear 
$g_{1T}^\perp(x,{\bf p}_T^2)$ can be accessed at leading-twist through the measurement of the DSA:

\vspace{-0.3cm}
\begin{equation}
2\langle{\rm cos}(\phi-\phi_S)\rangle^h_{L\perp}=2\frac{\int{d\phi d\phi_S {\rm cos}(\phi-\phi_S) \sigma_{LT}}}{\int{d\phi d\phi_S \sigma_{UU}}}=\frac{\mathit{C}\big[-\frac{{\bf P}_{h\perp}\cdot {\bf p}_T}{|{\bf P}_{h\perp}|~M}~g_{1T}^{\perp,q}(x,{\bf p}_T^2)~D_1^{q\rightarrow h}(z,{\bf K}_T^2) \big]}
{\mathit{C}\big[f_1^{~q}(x,{\bf p}_T^2)~D_1^{~q\rightarrow h}(z,{\bf K}_T^2) \big]}~,
\end{equation}

\noindent
where $\sigma_{LT}$ denotes the cross-section difference for opposite target polarization states, 
${\bf P}_{h\perp}$ is the transverse momentum of the produced hadron, $f_1^{~q}(x,{\bf p}_T^2)$ 
is the unpolarized distribution function and ${\mathit C}$ denotes a convolution integral over the 
intrinsic transverse momenta. Other Fourier components of $\sigma_{LT}$ are the sub-leading twist 
contributions  ${\rm cos}(\phi_S)$ and ${\rm cos}(2\phi-\phi_S)$, 
where the worm-gear $g_{1T}^\perp(x,{\bf p}_T^2)$ appears in convolution with the higher-twist
$\tilde{D}^\perp(z,{\bf K}_T^2)$ FF besides several other contributions of PDFs 
and FFs.

In this work, preliminary results for the Fourier components of the DSAs, measured at the HERMES 
experiment with a longitudinally polarized beam and transversely polarized protons, are discussed 
for identified pions and charged kaons. 

\section{Data analysis and results}

The data analysed was recorded during the 2003--2005 running period of the HERMES experiment using a 
longitudinally polarized $27.6$ GeV positron/electron beam and a transversely nuclear-polarized 
hydrogen gas target internal to the HERA storage ring at DESY. The open-ended target cell was fed by 
a polarized atomic-beam source~\cite{ref:Air05a} based on Stern-Gerlach separation and RF transitions of 
hyperfine states. The nuclear polarization of the atoms was flipped at $1-3$ minutes time intervals. 
Scattered leptons and any coincident hadrons were detected by the HERMES spectrometer~\cite{ref:Ack98}. 
Leptons are identified with an efficiency exceeding 98\% and a hadron contamination of less than 1\%. 
A dual-radiator RICH allows identification of the charged hadrons ($\pi,K,p$) in the $2-15$~GeV 
momentum range. Events were selected subject to the kinematic requirements $W^2>10$~GeV$^2$, 
$0.1<y<0.95$ and $Q^2>1$~GeV$^2$, where $W$ is the invariant mass of the photon-nucleon system, 
$y$ is the fractional beam energy transfered to the target and $-Q^2$ is the squared four-momentum of 
the virtual photon. Coincident hadrons were accepted in the range $0.2<z<0.7$ only. 

The three DSAs  $2\langle{\rm cos}(\phi-\phi_S)\rangle^h_{L\perp}$, 
$2\langle{\rm cos}(\phi_S)\rangle^h_{L\perp}$ and $2\langle{\rm cos}(\phi-2\phi_S)\rangle^h_{L\perp}$ were
extracted, together with six previously measured $A_{UT}$ SSAs~\cite{ref:Air10,ref:Air09,ref:Pap10}, 
in a maximum likelihood fit (unbinned in $\phi$ and $\phi_S$) of the selected SIDIS events, based on the 
probability density function:

\vspace{-0.4cm}
\begin{eqnarray} 
F(\phi,\phi_S)=1+P_T~\Big[2 \langle \sin(\phi+\phi_S) \rangle^h_{U\perp} ~\sin(\phi+\phi_S)~+... \Big]+ \nonumber\\
                   +P_T P_B~\Big[2 \langle \cos(\phi-\phi_S) \rangle^h_{L\perp} ~\cos(\phi-\phi_S)+ 2\langle \cos(\phi_S) \rangle^h_{L\perp}~\cos(\phi_S)+\nonumber\\
                   2\langle \cos(2\phi-\phi_S) \rangle^h_{L\perp} ~\cos(2\phi-\phi_S) \Big]~,
\label{PDF}
\end{eqnarray}

\noindent
where $P_T$ ($P_B$) denotes the target (beam) polarization and "..." stands for the contribution of 
the five $A_{UT}$ SSAs $2\langle \sin(\phi-\phi_S) \rangle^h_{U\perp}$, 
$2\langle \sin(3\phi-\phi_S) \rangle^h_{U\perp}$, 
$2\langle \sin(\phi_S) \rangle^h_{U\perp}$, $2\langle \sin(2\phi-\phi_S) \rangle^h_{U\perp}$, 
$2\langle \sin(2\phi+\phi_S) \rangle^h_{U\perp}$.

The systematic uncertainty, including contributions from acceptance effects, instrumental smearing, 
QED radiation and hadron misidentification, was evaluated as described in~\cite{ref:Air10}.
An additional $8.0 \%$ scale uncertainty, arising from the uncertainty on the beam and target 
polarization measurements, has to be considered.

\begin{figure}
  \includegraphics[height=.67\textheight]{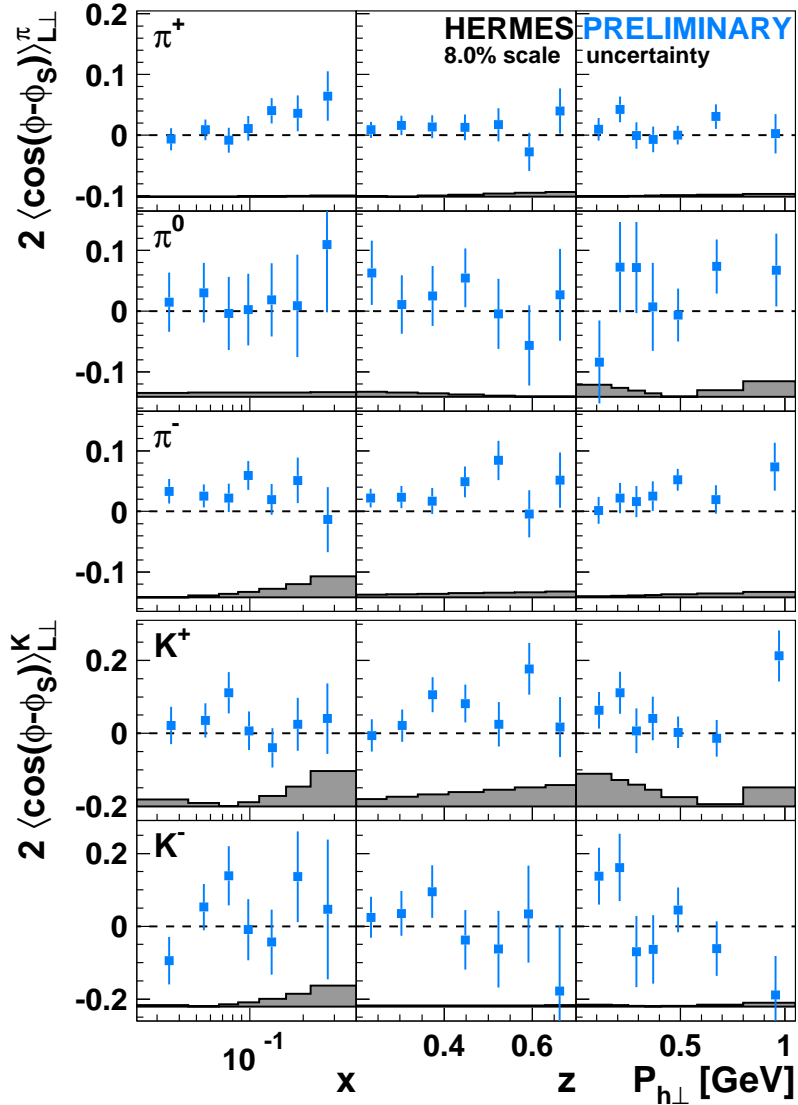}
  \caption{Preliminary results for the $2\langle \cos(\phi-\phi_S) \rangle^h_{L\perp}$ DSA
   amplitudes for pions and charged kaons as a function of $x$, $z$ or $P_{h\perp}$.
   The shaded bands represent the systematic uncertainty. A common $8.0 \%$ scale 
   uncertainty arises from the precision of the beam and target polarization measurements.}
\label{results}
\end{figure}

The preliminary results for the $2\langle \cos(\phi-\phi_S) \rangle^h_{L\perp}$ asymmetry amplitudes are 
reported in Fig.~\ref{results} for pions and charged kaons as a function of $x$, $z$ or $P_{h\perp}$.
The results show a positive amplitude for $\pi^-$  and a hint of a positive signal also for $\pi^+$
and $K^+$, whereas amplitudes consistent with zero are observed for $\pi^0$ and $K^-$. 
The positive amplitude for $\pi^-$ reported here is similar to that recently 
measured at Jefferson Lab (E06010 experiment in Hall-A) but on a transversely polarized $^3He$ 
(i.e. neutron) target~\cite{ref:Hua11-Jia10}. The amplitudes for the sub-leading twist DSAs 
$2\langle{\rm cos}(\phi_S)\rangle^h_{L\perp}$ and $2\langle{\rm cos}(2\phi-\phi_S)\rangle^h_{L\perp}$, 
not shown, are both consistent with zero for all measured mesons.

\bibliographystyle{aipproc}   

\bibliography{sample}

\IfFileExists{\jobname.bbl}{}
 {\typeout{}
  \typeout{******************************************}
  \typeout{** Please run "bibtex \jobname" to optain}
  \typeout{** the bibliography and then re-run LaTeX}
  \typeout{** twice to fix the references!}
  \typeout{******************************************}
  \typeout{}
 }

\end{document}